\documentclass[aps,twocolumn]{revtex4-1}
\usepackage[dvips]{graphicx}
\usepackage{amsmath}
\usepackage{amssymb}
\usepackage{bm}

\begin{document}
\title{Machine-Learning Study using Improved Correlation Configuration \\
and Application to Quantum Monte Carlo Simulation} 

\author{Yusuke Tomita$^{1}$}
\email{ytomita@shibaura-it.ac.jp}
\author{Kenta Shiina$^{2,3}$}
\email{16879316kenta@gmail.com}
\author{Yutaka Okabe$^{2}$}
\email{okabe@phys.se.tmu.ac.jp}
\author{Hwee Kuan Lee$^{3,4,5,6}$}
\email{leehk@bii.a-star.edu.sg}
\affiliation{
$^1$College of Engineering, Shibaura Institute of Technology, 
Saitama 330-8570, Japan \\
$^2$Department of Physics, Tokyo Metropolitan University, 
Hachioji, Tokyo 192-0397, Japan \\
$^3$Bioinformatics Institute, 
Agency for Science, Technology and Research (A*STAR),
30 Biopolis Street, \#07-01 Matrix, 
138671, Singapore \\
$^4$School of Computing, National University of Singapore, 
13 Computing Drive, 
117417, Singapore \\
$^5$Singapore Eye Research Institute (SERI), 
11 Third Hospital Ave, 
168751, Singapore \\
$^6$Image and Pervasive Access Laboratory (IPAL), 
1 Fusionopolis Way, \#21-01 Connexis (South Tower), 
138632, Singapore
}

\date{\today}

\begin{abstract}
We use the Fortuin-Kasteleyn representation based improved estimator of the correlation configuration 
as an alternative to the ordinary correlation configuration 
in the machine-learning study of the phase classification 
of spin models.
The phases of classical spin models are classified 
using the improved estimators, 
and the method is also applied to the quantum 
Monte Carlo simulation using the loop algorithm. 
We analyze the Berezinskii-Kosterlitz-Thouless (BKT) transition 
of the spin 1/2 quantum XY model on the square lattice. 
We classify the BKT phase and the paramagnetic phase of the 
quantum XY model using the machine-learning approach. 
We show that the classification of the quantum XY model 
can be performed by using the training data of 
the classical XY model. 
\end{abstract}

\maketitle

Remarkable developments of machine-learning based techniques 
have been made in the past decade, which have given an impact 
on many areas in industry including automated driving, 
healthcare, etc. At the same time, the potential of 
machine learning for fundamental research has gained 
increasing interest. Statistical physics is one of 
such scientific disciplines~\cite{Carleo}. 

Carrasquilla and Melko~\cite{Carrasquilla} used a technique 
of supervised learning to propose a paradigm that is complementary 
to the conventional approach of studying interacting spin systems. 
By using large data sets of spin configurations, 
they classified and identified a high-temperature paramagnetic phase 
and a low-temperature ferromagnetic phase. 
It was similar to image classification using 
machine learning. 
They demonstrated the use of neural networks for the study 
of the two-dimensional (2D) Ising model and 
an Ising lattice gauge theory. 

Shiina \textit{et al.}~\cite{Shiina} reported a machine-learning study 
on phase transitions.  
The configuration of a long-range spatial correlation 
was treated instead of the spin configuration itself. 
By doing so, a similar treatment was provided to various spin models 
including the multi-component systems and the systems 
with a vector order parameter. 
Not only the second-order and the first-order transitions 
but also the Berezinskii-Kosterlitz-Thouless (BKT) 
transition~\cite{Berezinskii1,Berezinskii2,kosterlitz,kosterlitz2} 
was studied. 
The disordered and the ordered phases, along with the BKT type 
topological phase, were successfully classified.

Cluster algorithms~\cite{sw87,Wolff89} have been used
to overcome slow dynamics in the Monte Carlo simulation.
Swendsen and Wang (SW)~\cite{sw87} applied the
Fortuin-Kasteleyn (FK)~\cite{KF,FK} representation to identify 
clusters of spins. 
The single-cluster variant of the cluster algorithm was proposed 
by Wolff~\cite{Wolff89}. 
Wolff also proposed the idea of the embedded cluster 
formalism~\cite{Wolff89,Wolff89b,Wolff90} to treat vector spin models. 
By projecting a vector spin onto a randomly chosen unit vector, 
the Ising degrees of freedom are picked up.  Then, the cluster spin flip 
can be performed with the FK cluster. 
A further advantage of cluster algorithms is that they lead to 
so-called improved estimators~\cite{Wolff90} which are designed 
to reduce the statistical errors.  
In calculating spin correlations, only the spin pair  
belonging to the same FK cluster should be considered.  
The feature of manifesting the spin correlations in a spin configuration
is utilized in the probability-changing cluster algorithm, 
a self-adapted algorithm to tune the critical point 
automatically~\cite{tomita2001}.

Evertz \textit{et al.}~\cite{Evertz93,Evertz03} presented 
another type of cluster algorithm, which is called loop algorithm. 
In treating vertex models, closed paths of bonds are flipped. 
Constraints at the vertices are automatically satisfied. 
The loop algorithm was applied to quantum spin systems 
in the worldline representation~\cite{Beard,Kawashima,Gubernatis}. 
The improvements accomplished on the quantum Monte 
Carlo simulation was largely due to 
the global update, in which configurations are updated in
units of some non-local clusters. 
By using the loop algorithm, non-diagonal quantities
can be measured. 

In this study, we consider the improved estimator for the correlation 
configuration in the cluster representation. 
We use the machine-learning method of 
Shiina \textit{et al.}~\cite{Shiina} for the classification 
of phases using the improved correlation configuration. 
Then, we apply this technique to quantum spin systems. 
As an example, we show the results of the spin 1/2 XY model 
on the square lattice.  This model exhibits the BKT 
transition~\cite{Harada}. 

\begin{figure}
\centering

\begin{minipage}{2.8cm}
\begin{flushleft}
{\textbf{(a)}} \ spin
\end{flushleft}
\centering
\includegraphics[width=2.7cm,clip]{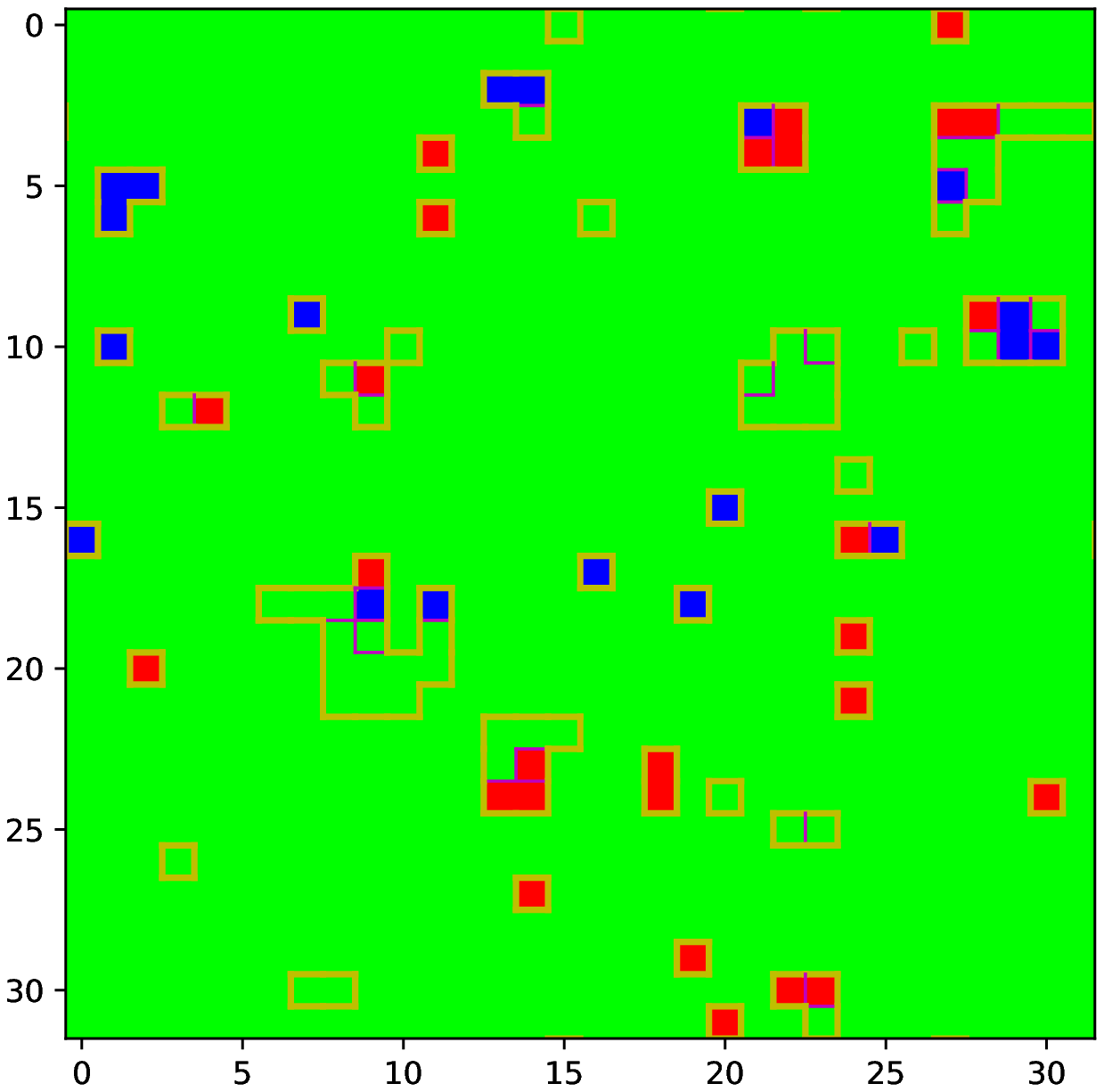}
\end{minipage}
\begin{minipage}{2.8cm}
\vspace*{4mm}
\begin{flushleft}
{\textbf{(b)}} \ correlation
\end{flushleft}
\centering
\includegraphics[width=2.7cm,clip]{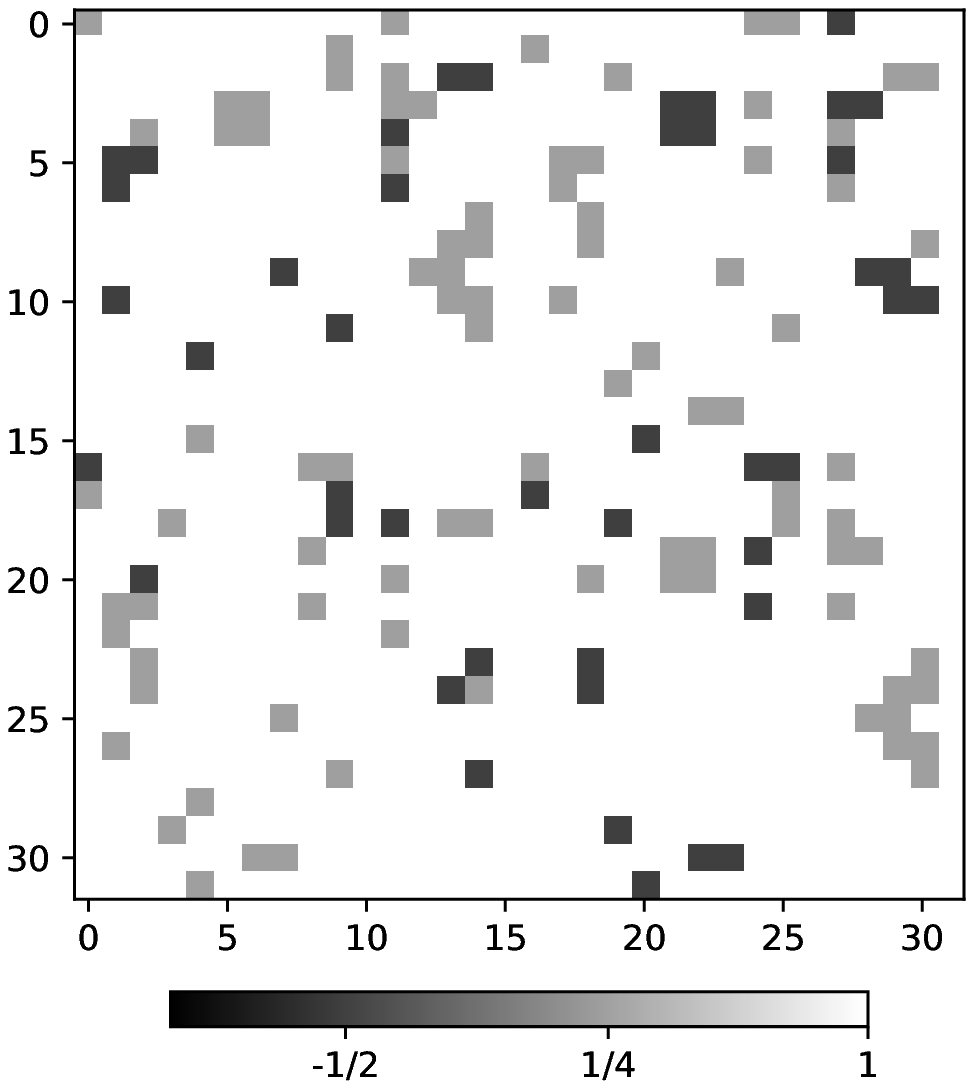}
\end{minipage}
\begin{minipage}{2.8cm}
\vspace*{4mm}
\begin{flushleft}
{\textbf{(c)}} \ improved
\end{flushleft}
\centering
\includegraphics[width=2.7cm,clip]{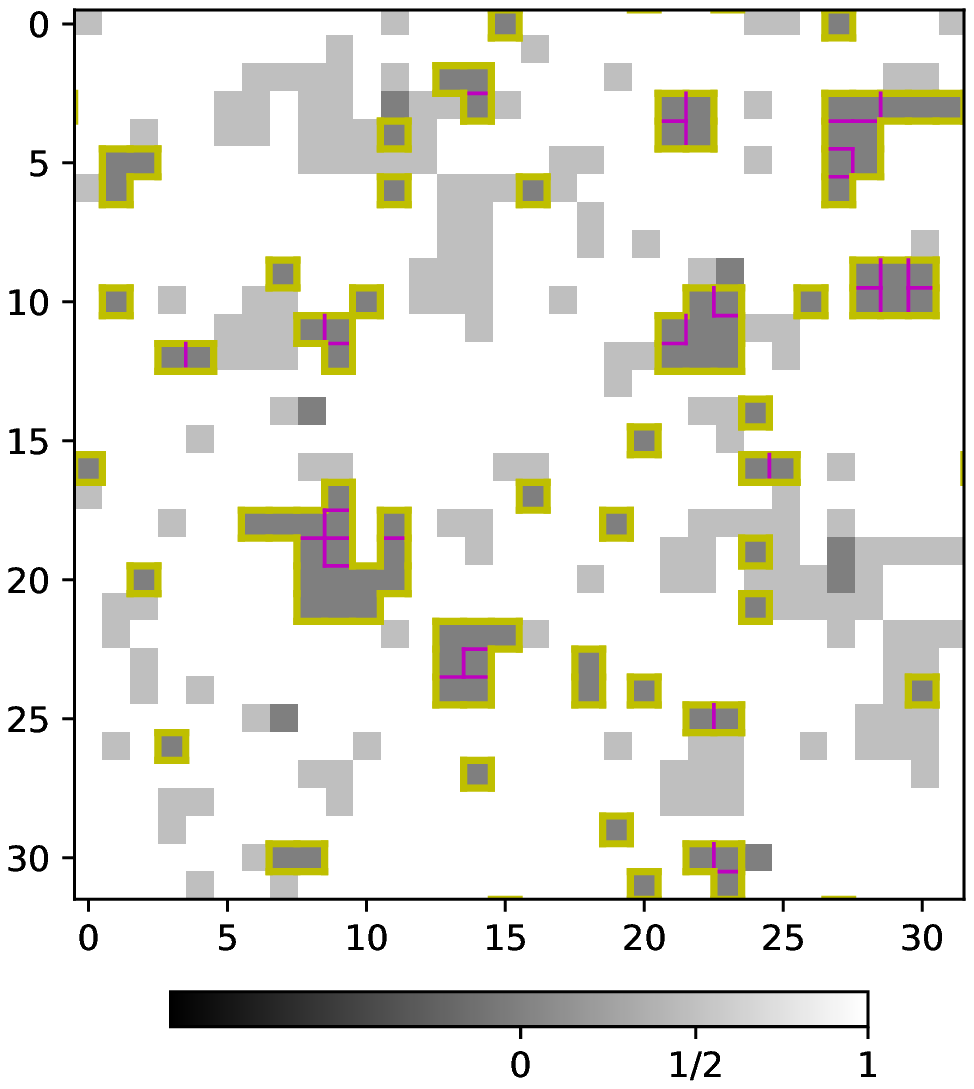}
\end{minipage}

\begin{minipage}{2.8cm}
\begin{flushleft}
{\textbf{(d)}} \ spin
\end{flushleft}
\centering
\includegraphics[width=2.7cm,clip]{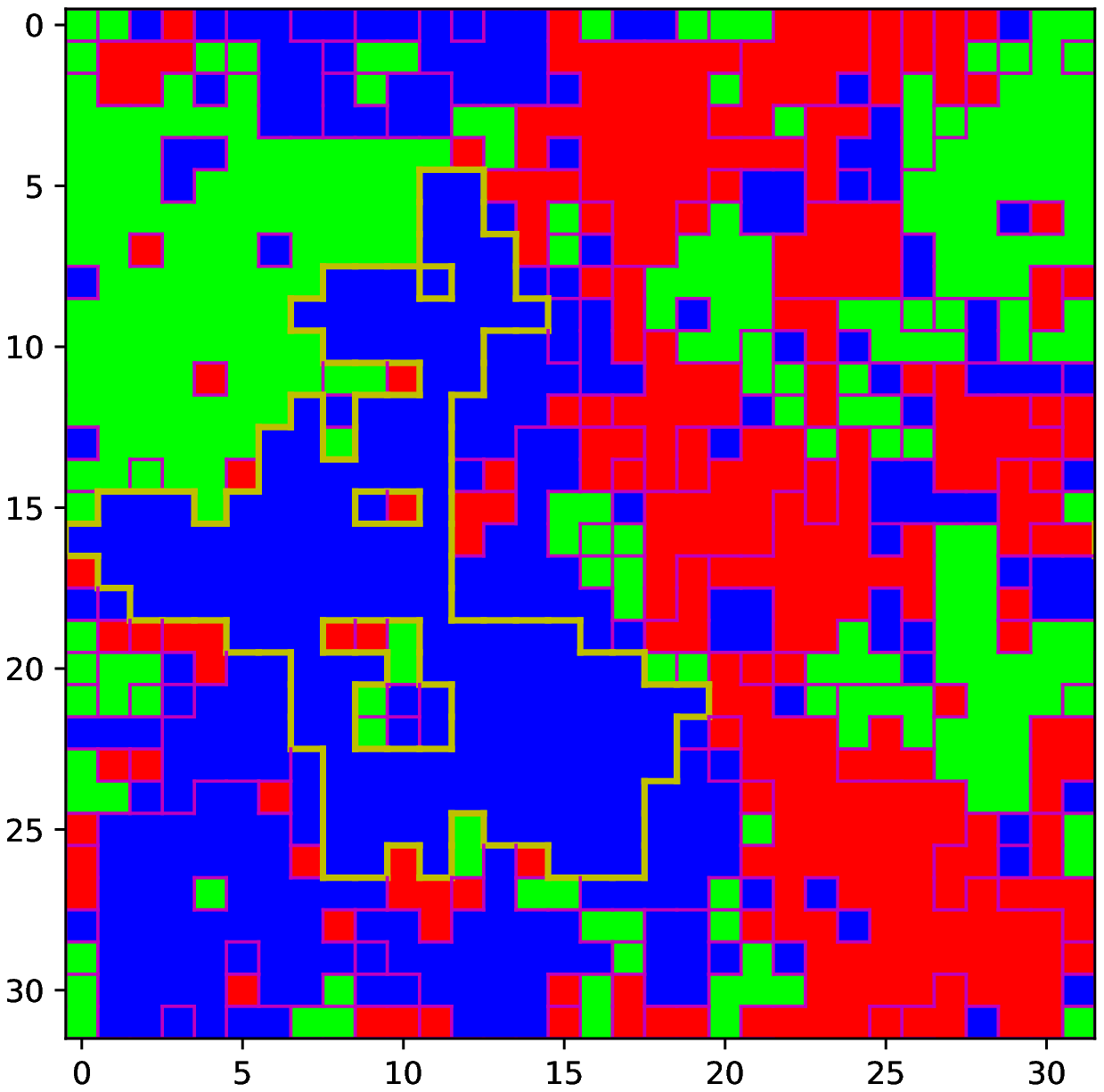}
\end{minipage}
\begin{minipage}{2.8cm}
\vspace*{4mm}
\begin{flushleft}
{\textbf{(e)}} \ correlation
\end{flushleft}
\centering
\includegraphics[width=2.7cm,clip]{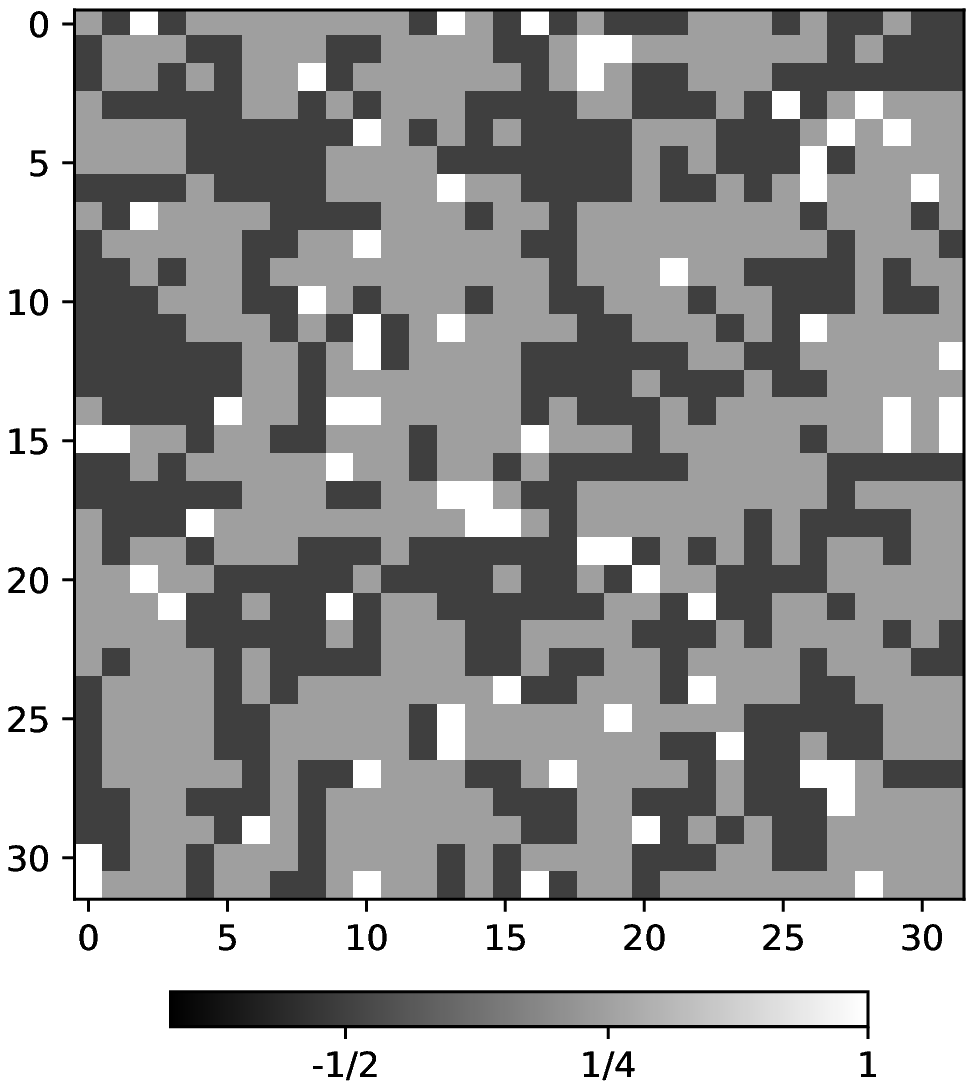}
\end{minipage}
\begin{minipage}{2.8cm}
\vspace*{4mm}
\begin{flushleft}
{\textbf{(f)}} \ improved
\end{flushleft}
\centering
\includegraphics[width=2.7cm,clip]{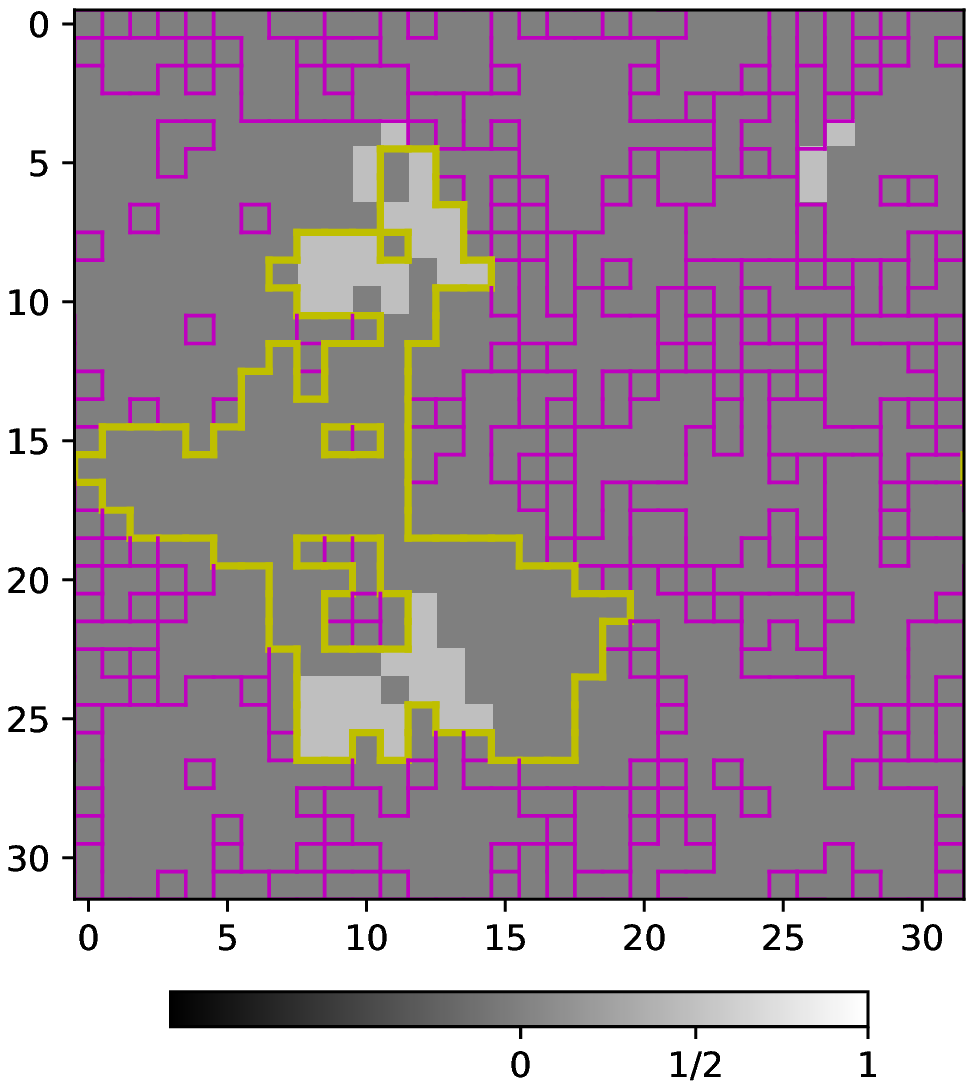}
\end{minipage}

\caption{
Examples of the spin configuration $\{ s_i \}$ ((a), (d)), 
correlation configuration $\{ g_i(L/2) \}$ ((b), (e)), and
improved correlation configuration $\{ \tilde{g}_i(L/2) \}$ ((c), (f)) 
of the 2D 3-state Potts model. 
The upper figures ((a)-(c)) are snapshots at the low temperature, $T=0.9$, 
and the lower figures ((d)-(f)) are those at the high temperature, $T=1.06$. 
The borders of FK clusters for spin configuration are drawn by lines. 
They are copied in improved correlation configuration. 
}
\label{fig:config_Potts}
\end{figure}

We consider the configuration of the spin correlation 
with the distance of the half of the system size, $L/2$.  
We note that this type of correlation function was used 
along with the generalized scheme for the probability-changing 
cluster algorithm~\cite{tomita2002b}. 
For the $q$-state Potts model (including the Ising model), 
the correlation between two spins becomes 1 for the same spin pair, 
whereas it becomes $-1/(q-1)$ for the pair of different states. 
In the improved estimator for the cluster representation, 
the correlation becomes 1 for the spin pair belonging to 
the same FK cluster, whereas it becomes 0 for the spins 
of different clusters. 
When the embedded algorithm for the continuous spins is used, 
the projection of spins onto a randomly chosen reflection axis is made. 
We denote the site-dependent correlation configuration 
as $g_i(L/2)$. 
For actual calculation, we treat the average value 
of the $x$-direction and the $y$-direction, that is, 
\begin{equation}
  g_i(L/2) = (g[s_{x_i,y_i},s_{x_i+L/2,y_i}]+g[s_{x_i,y_i},s_{x_i,y_i+L/2}])/2,
\label{g_i}
\end{equation}
where $g[s, s']$ denotes a spin-spin correlation 
between a spin pair $s$ and $s'$.

In Fig.~\ref{fig:config_Potts}, we show the examples of
the spin configuration $\{ s_i \}$, 
correlation configuration $\{ g_i(L/2) \}$, and 
improved correlation configuration $\{ \tilde{g}_i(L/2) \}$ 
of the 2D 3-state Potts model. The spin configuration 
is generated by the Monte Carlo simulation, and the correlation 
configuration and the improved correlation configurations 
are calculated from the spin configuration. 
The upper figures are snapshots at the low temperature, $T=0.9$, 
and the lower figures are those at the high temperature, $T=1.06$. 
Temperatures are measured in the unit of the coupling $J$.
We note that the exact second-order transition temperature $T_c$ 
for this model is known as 
$1/\ln(1+\sqrt{3}) \approx 0.995$.

Spins are displayed in one of three colors, red, green, or blue. 
The ordinary correlation takes a value of $1, -1/2$, or $+1/4$, 
whereas the improved correlation takes a value of $1, 0$, or $+1/2$. 
The both of correlations from $+1$ to $-1$ are mapped in gray scale 
from 255 (white) to 0 (black). 
The permutation of three-state spins yields an essentially identical
configuration, and the correlation configurations are 
invariant under the permutation. 
The borders of FK clusters for spin configuration 
are drawn by lines in Figs.~\ref{fig:config_Potts}(a) and (d). 
They are copied in improved correlation configuration. 
The border of the largest cluster is drawn by yellow thick line 
for convenience. 

At high temperatures, the spin configurations and the correlation 
configurations are randomly distributed, and the fluctuation of 
these quantities gives the susceptibility. 
In the improved correlation, the cancellation among 
different FK clusters are automatically satisfied. 
Figure~\ref{fig:config_Potts}(e)
and \ref{fig:config_Potts}(f)
show the difference between the two correlation configurations.
While the ordinary correlation configuration in Fig.~\ref{fig:config_Potts}(e)
fluctuates in space,
a couple of brighter areas in the largest cluster
show the improved correlation in Fig.~\ref{fig:config_Potts}(f).

As a supplemental material, we provide animations 
of the spin configuration, the correlation configuration, 
and the improved correlation configuration for the 2D Ising model 
(mp4 files) for convenience~\cite{Supple}. 
The animations at various temperatures are compared 
at the low temperature ($T=2.1$), at $T_c=2.269$, 
and at the high temperature ($T=2.4$). 
The system sizes are $L$=32 and 64.

We use the same technique of supervised learning 
as Shiina \textit{et al.}~\cite{Shiina} 
for the classification of the phases of spin systems. 
We consider a fully connected neural network implemented 
with a standard TensorFlow library~\cite{TF} using the 100-hidden 
unit model to classify the ordered, the BKT,  and the disordered phases. 
For the input layer, we use the improved correlation 
configurations $\{ \tilde{g}_i(L/2) \}$. 
We have used a cross-entropy cost function supplemented with 
an $L2$ regularization term. 
The neural networks were trained using the Adam method~\cite{Adam}. 

\begin{figure}[t]
\centering
\begin{flushleft}
\quad\quad{\textbf{(a)}}
\end{flushleft}
\includegraphics[width=7.2cm,clip]{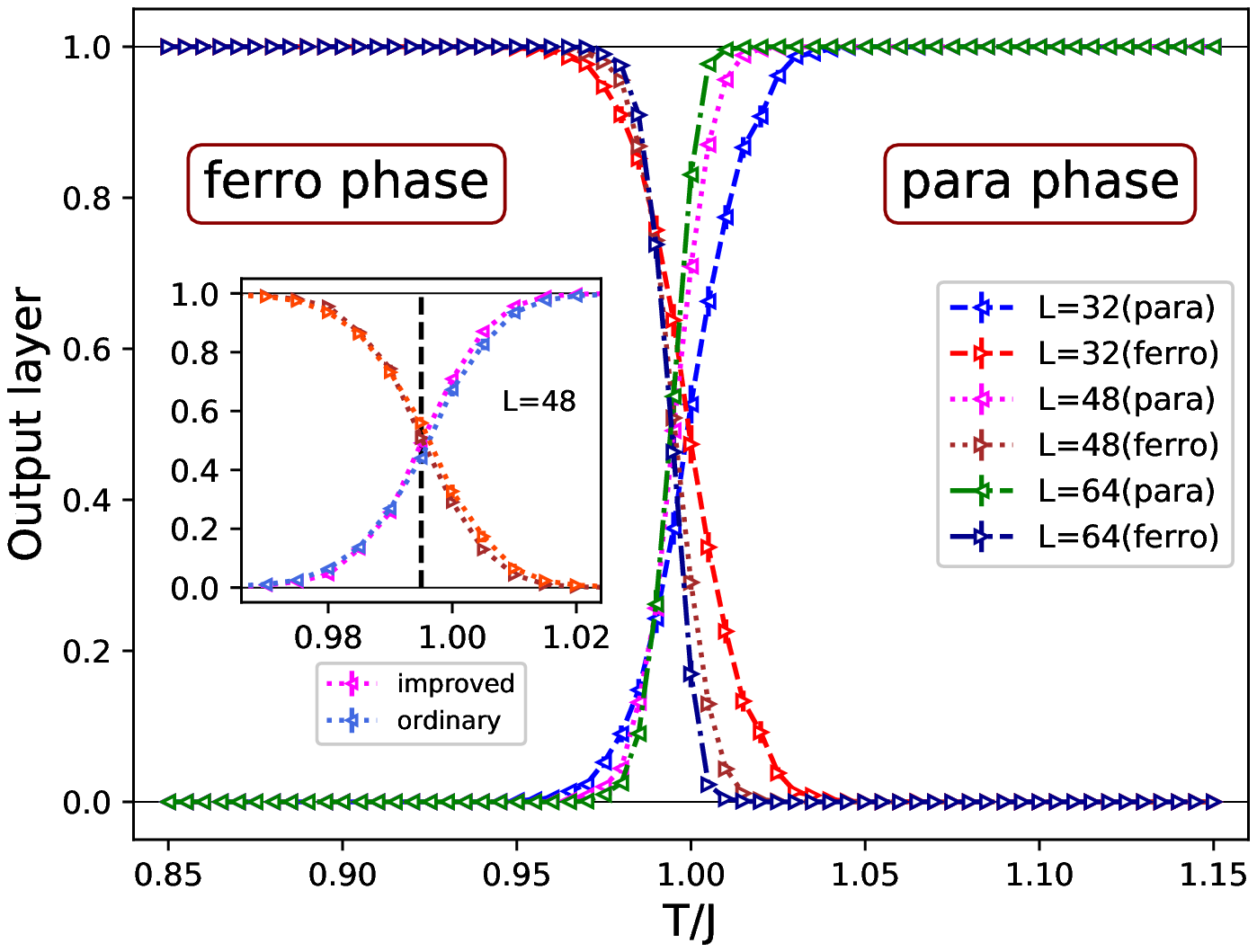}
\begin{flushleft}
\quad\quad{\textbf{(b)}}
\end{flushleft}
\includegraphics[width=7.2cm,clip]{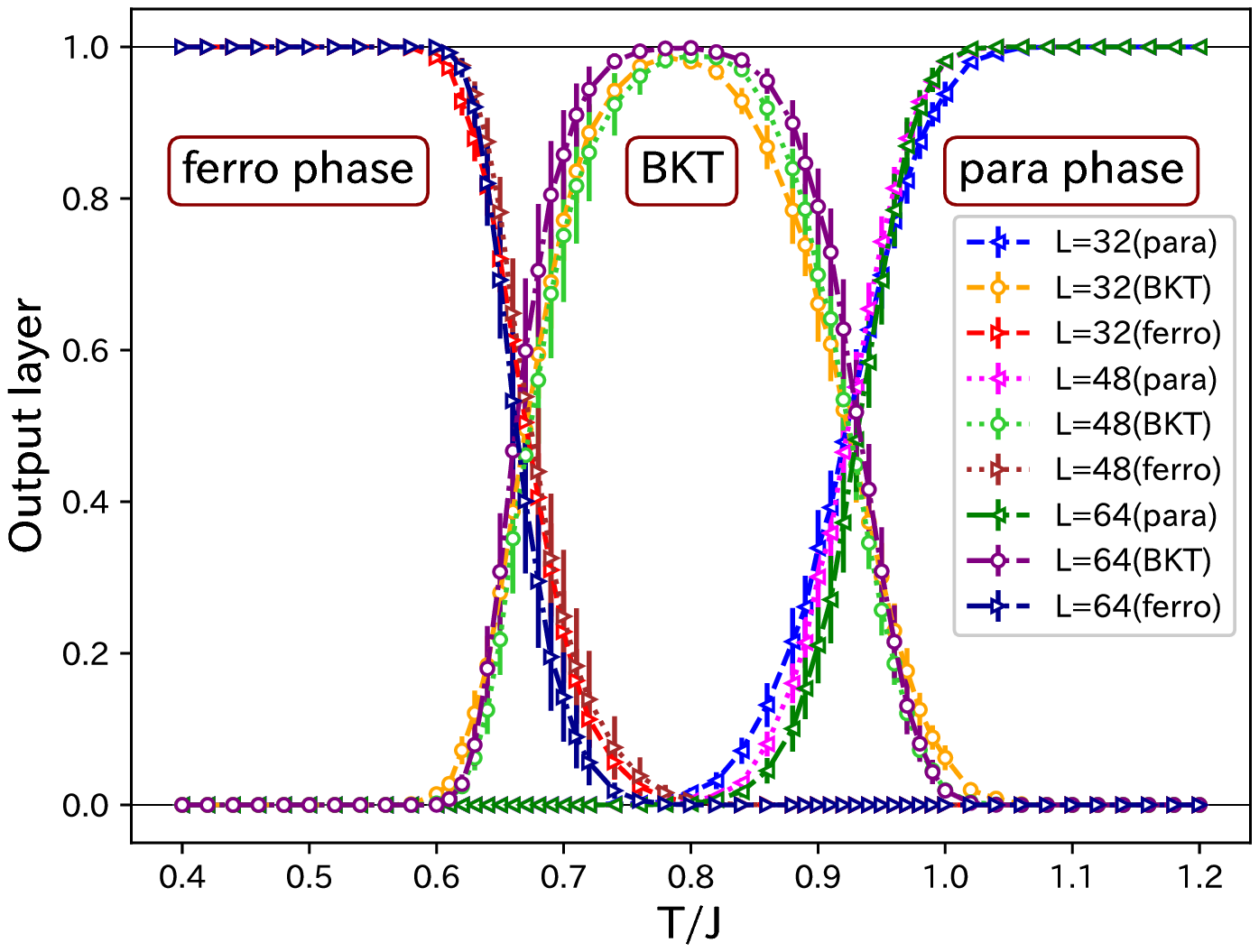}
\caption{
(a) The output layer averaged over a test set as a function of $T$ 
for the 2D 3-state Potts model. The system sizes are $L$ = 
32, 48, and 64. 
The samples of $T$ within the ranges $0.85 \le T \le 0.94$ and 
$1.06 \le T \le 1.15$ are used for the training data. 
In the inset, the comparison is made between the results of 
improved correlation and those of ordinary correlation 
in the case of $L=48$. The exact $T_c$, $1/\ln(1+\sqrt{3}) 
\approx 0.995$, is shown by dashed line 
for convenience. 
(b) The same plot for the 2D 6-state clock model. 
The system sizes are the same. 
The samples of $T$ within the ranges $0.4 \le T \le 0.64$, 
$0.77 \le T \le 0.83$, and 
$0.96 \le T \le 1.2$ are used for the training data. 
}
\label{fig:classical}
\end{figure}

We first analyzed the 2D 3-state Potts model. 
The output layer averaged over a test set as a function of $T$ 
for the 2D 3-state Potts model is shown in Fig.~\ref{fig:classical}(a). 
The probabilities of predicting the phases, the disordered or the ordered, 
are plotted for each temperature. 
The system sizes are $L$ = 32, 48, and 64. 
The samples of $T$ within the ranges $0.85 \le T \le 0.94$ and 
$1.06 \le T \le 1.15$ were used for the training data. 
We have not used the samples close to $T_c$ for the training data 
because we assumed the situation that the exact $T_c$ is not known. 
For a whole temperature range, around 35,000 training data sets 
are used, and we use 500 test data sets for each temperature. 
Ten independent calculations were 
performed to provide error analysis. 
This figure corresponds to Fig.~2(a) of Ref.~\cite{Shiina},
and we again observe that the neural network could successfully classify 
the disordered and ordered phases using the improved correlation 
configuration. 
In the inset of Fig.~\ref{fig:classical}(a), we show the comparison 
of the results of improved correlation (the present study) 
and those of the previous study~\cite{Shiina} 
of ordinary correlation in the case of $L=48$. 
We used the same conditions for both training data and 
test data of improved and ordinary correlations 
produced from the same spin configurations. 
The point that the probabilities of predicting two phases are 50\% 
is slightly more close to the exact critical temperature, shown in 
dashed line in the inset, for the improved correlation, 
but the difference is small.  The advantage of the improved estimator 
appears at high enough temperatures (compare 
Fig.~\ref{fig:config_Potts}(f) with \ref{fig:config_Potts}(e)). 

We next consider the 2D 6-state clock model. 
Because of the discreteness, there are two transitions. 
One is a higher BKT transition, $T_2$, between the disordered 
and BKT phases, and the other is a lower transition, $T_1$, 
between the BKT and ordered phases. 
The output layer averaged over a test set as a function of $T$ 
for the 2D 6-state clock model is shown in Fig.~\ref{fig:classical}(b). 
The system sizes are $L$ = 32, 48, and 64. 
The samples of $T$ within the ranges $0.4 \le T \le 0.64$, 
$0.77 \le T \le 0.83$, and 
$0.96 \le T \le 1.2$ were used for the low-temperature, 
mid-range temperature, and high-temperature training data, respectively. 
The recent numerical estimates of $T_1$ and $T_2$ for the 
6-state clock model are 0.701(5) and 0.898(5), respectively~\cite{Surungan}. 
This figure corresponds to Fig.~4(a) of Ref.~\cite{Shiina},
and the present figure again shows the successful classification 
into the three phases. 

We have classified the phases of transitions 
by means of the machine-learning approach by 
Shiina \textit{et al.}~\cite{Shiina} 
using improved correlation configuration.  
There is no appreciable difference of accuracy 
between the use of the correlation configuration 
and that of the improved correlation configuration. 
The result indicates that the machine-learning based phase 
classification is robust; 
that is, the phase classification 
does not discriminate the improved correlation configuration 
from the ordinary one.

Many applications of the loop updating method have been done 
for quantum systems. 
Here, we consider the quantum spin 1/2 XY model in two dimensions, 
which clearly demonstrated the utility of the loop 
algorithm~\cite{Harada}. 
The Hamiltonian is written as
\begin{equation}
 H = -J \sum_{\langle i,j \rangle}
      (\hat{S}_i^{x}\hat{S}_j^{x} + \hat{S}_i^{y}\hat{S}_j^{y}).
\end{equation}
Here, the spin operators $\hat{S}^{x,y}$ are one-half of 
the Pauli matrices $\sigma^{x,y}$. 
The summation is taken over the nearest-neighbor pairs. 
This model exhibits the BKT transition 
at around $T=0.342$~\cite{Harada}. 

We performed the quantum Monte Carlo simulation using the loop 
algorithm, and calculated the spatial correlation with the distance of $L/2$. 
A $D$-dimensional quantum system can be treated as a 
$(D+1)$-dimensional classical system with an extra dimension 
of imaginary time. In calculating the spatial correlation, 
the summation over the imaginary-time axis is taken. 
The $\hat{S}^x$-component of the correlation function is 
calculated as~\cite{Brower}
\begin{align}
  \tilde{g}^x_{\bm{r}_i,\bm{r}_j}
  &= \frac{4}{\beta^2} \int_{0}^{\beta} \int_{0}^{\beta}
  \hat{S}^{x}(\bm{r}_i,\tau_1)\hat{S}^{x}(\bm{r}_j,\tau_2) 
  \ d\tau_1 d\tau_2
  \nonumber \\
  &= \frac{1}{\beta^2} \int_{0}^{\beta} \int_{0}^{\beta}
  \delta_\ell(\bm{r}_i,\tau_1;\bm{r}_j,\tau_2) 
  \ d\tau_1 d\tau_2,
\end{align}
where $\delta_\ell(\cdot)$ is the function that returns 1 if 
the loop of the position $\bm{r}_i$ and the time $\tau_1$ and 
that of the position $\bm{r}_j$ and the time $\tau_2$ belong to the same loop, 
whereas returns 0 otherwise.
Due to the O(2) symmetry of the model,
the $\hat{S}^y$-component of the correlation function
is exactly the same as the $\hat{S}^x$-component~\cite{Brower}.
The factor 4 is introduced for the comparison of the spin one-half system 
with the classical model. 
We checked our calculation by the consistency with the precise 
calculations at $T=0$~\cite{Sandvik,Lin}.

\begin{figure}[t]
\centering
  \begin{flushleft} 
  (a) quantum \hspace*{2.0mm} $T=0.24$ \hspace*{9.0mm} 
  (b) quantum \hspace{2.0mm} $T=0.44$ 
  \end{flushleft}
\includegraphics[width=3.5cm,clip]{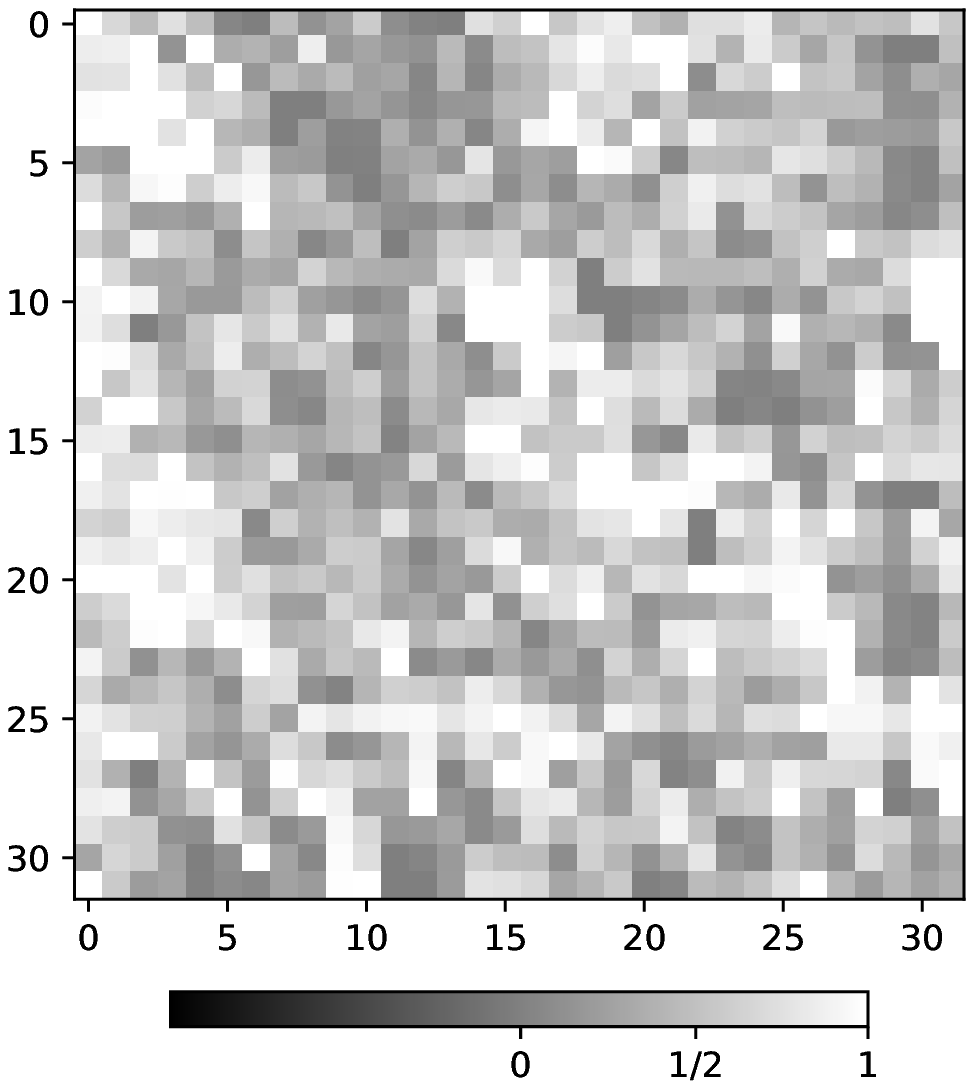}
\hspace*{4mm}
\includegraphics[width=3.5cm,clip]{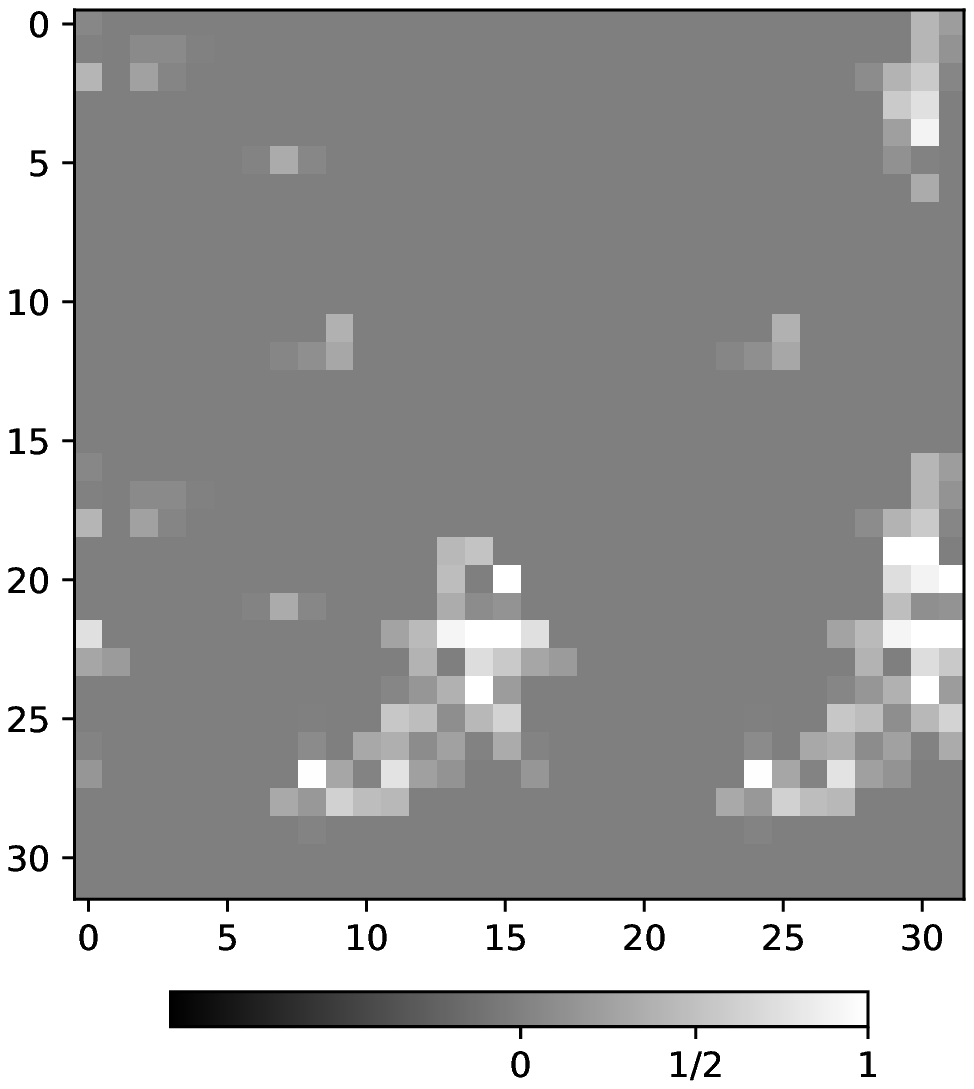}
  \begin{flushleft} 
  (c) classical \hspace*{2.0mm} $T=0.6$ \hspace*{11.0mm} 
  (d) classical \hspace{2.0mm}$T=1.2$
  \end{flushleft}
\includegraphics[width=3.5cm,clip]{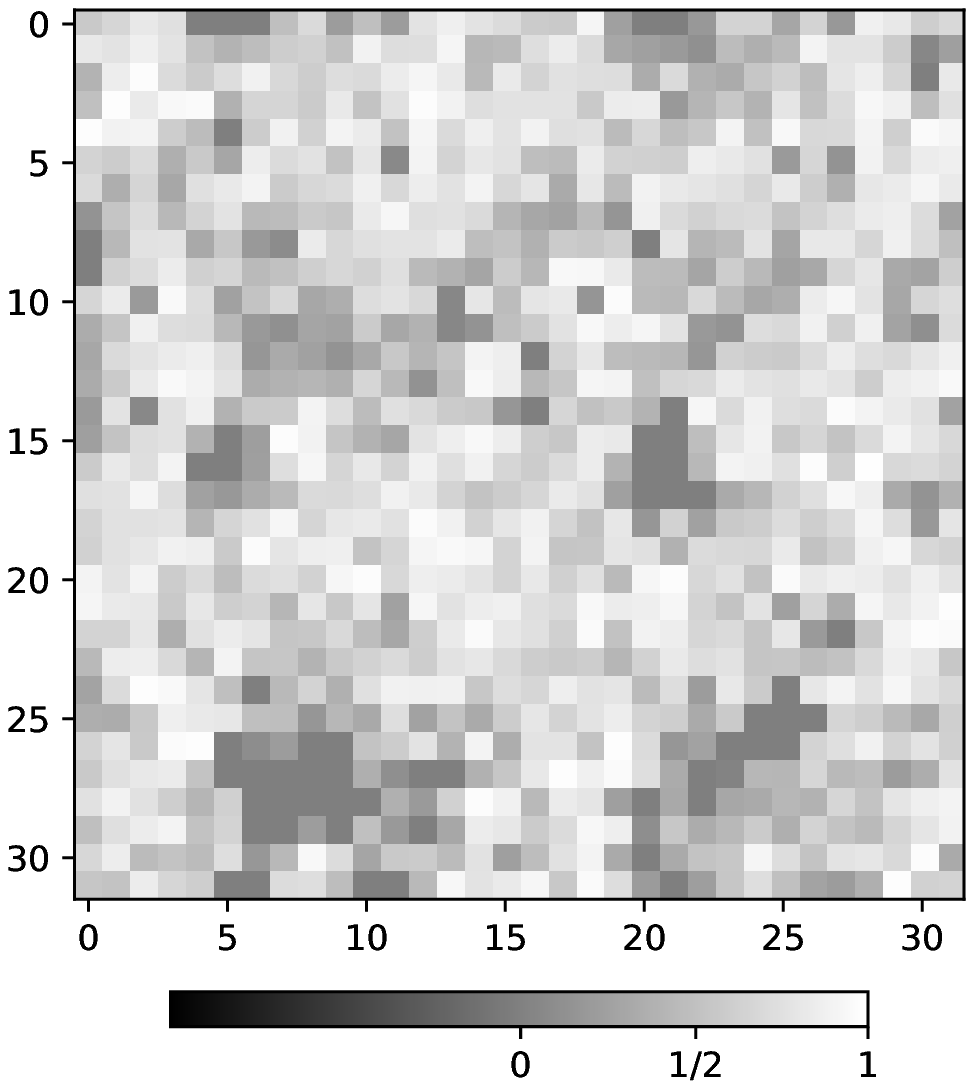}
\hspace*{4mm}
\includegraphics[width=3.5cm,clip]{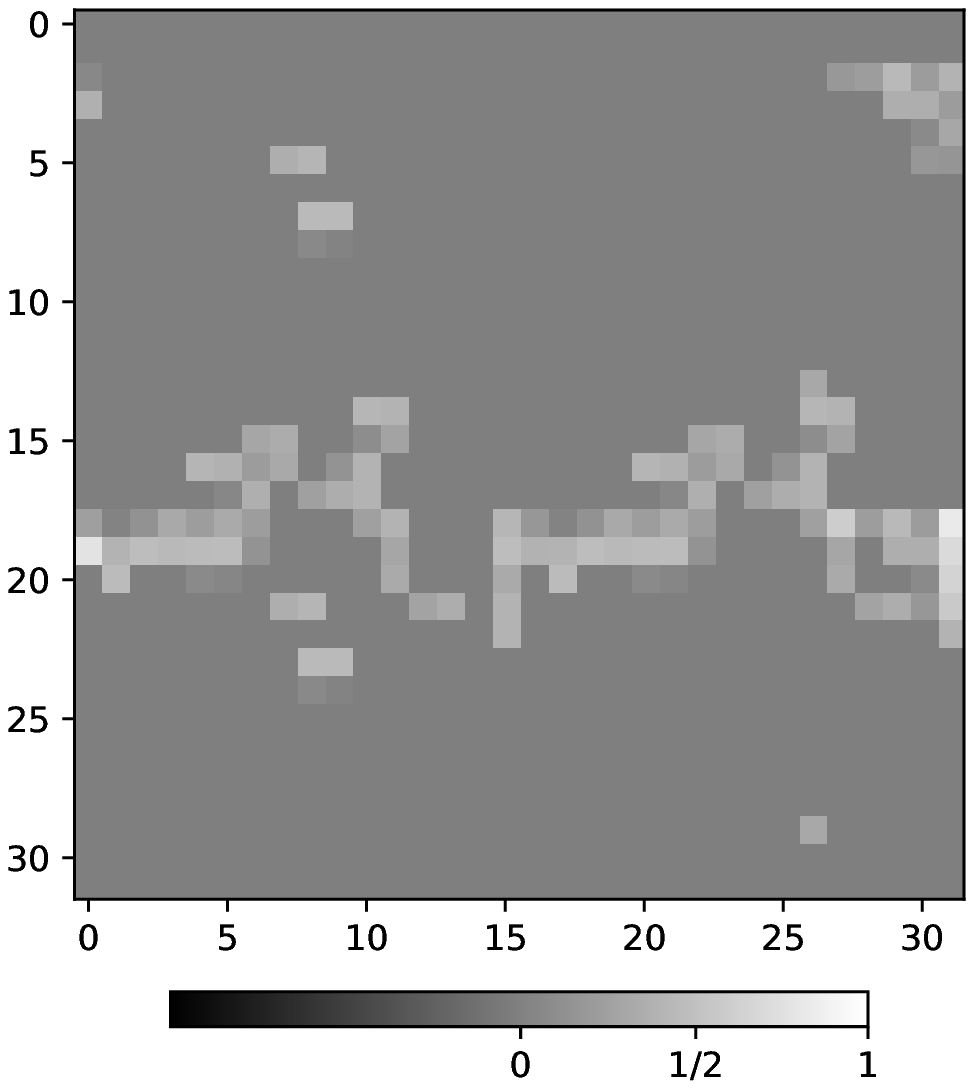}
\caption{
Comparison of snapshots of $\{\tilde{g}_i(L/2)\}$ of the 2D 
quantum ((a), (b)) and classical ((c), (d)) XY models. 
Examples of snapshots below $T_{\mathrm{BKT}}$ ((a), (c)) and 
those above $T_{\mathrm{BKT}}$ ((b), (d)) are displayed.
}
\label{fig:snap_xy}
\end{figure}

It is instructive to compare the correlation configurations 
of the quantum XY model and the classical XY model. 
In Fig.~\ref{fig:snap_xy}, examples of the snapshots of 
$\{\tilde{g}_i(L/2)\}$ of two models are displayed.  
At high temperatures above $T_{\mathrm{BKT}}$ ((b), (d)), both improved 
configurations represent the behavior of finite correlation length. 
At low temperatures below $T_{\mathrm{BKT}}$ ((a), (c)), 
they are different from the high-temperature configurations 
and at the same time they are different from the behavior of 
the ordered state, which was shown in Fig.~\ref{fig:config_Potts}(c). 
(Note that the precise estimate of the BKT temperature of the classical 
XY model is $T_{\mathrm{BKT}}$ = 0.8929~\cite{Hasenbusch}.)

\begin{figure}[t]
\centering
\begin{flushleft}
\quad\quad{\textbf{(a)}}
\end{flushleft}
\includegraphics[width=7.2cm,clip]{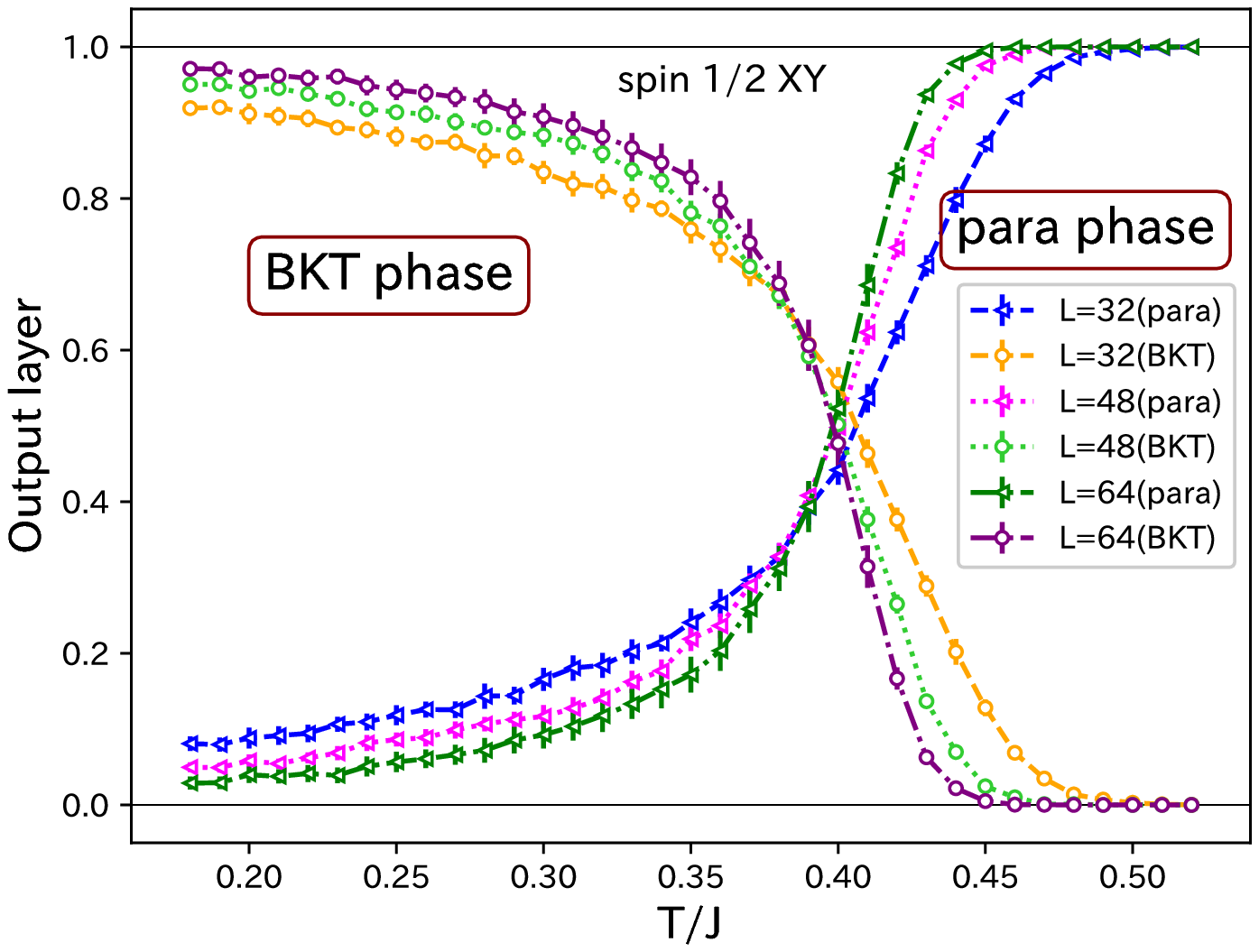}
\begin{flushleft}
\quad\quad{\textbf{(b)}}
\end{flushleft}
\includegraphics[width=7.2cm,clip]{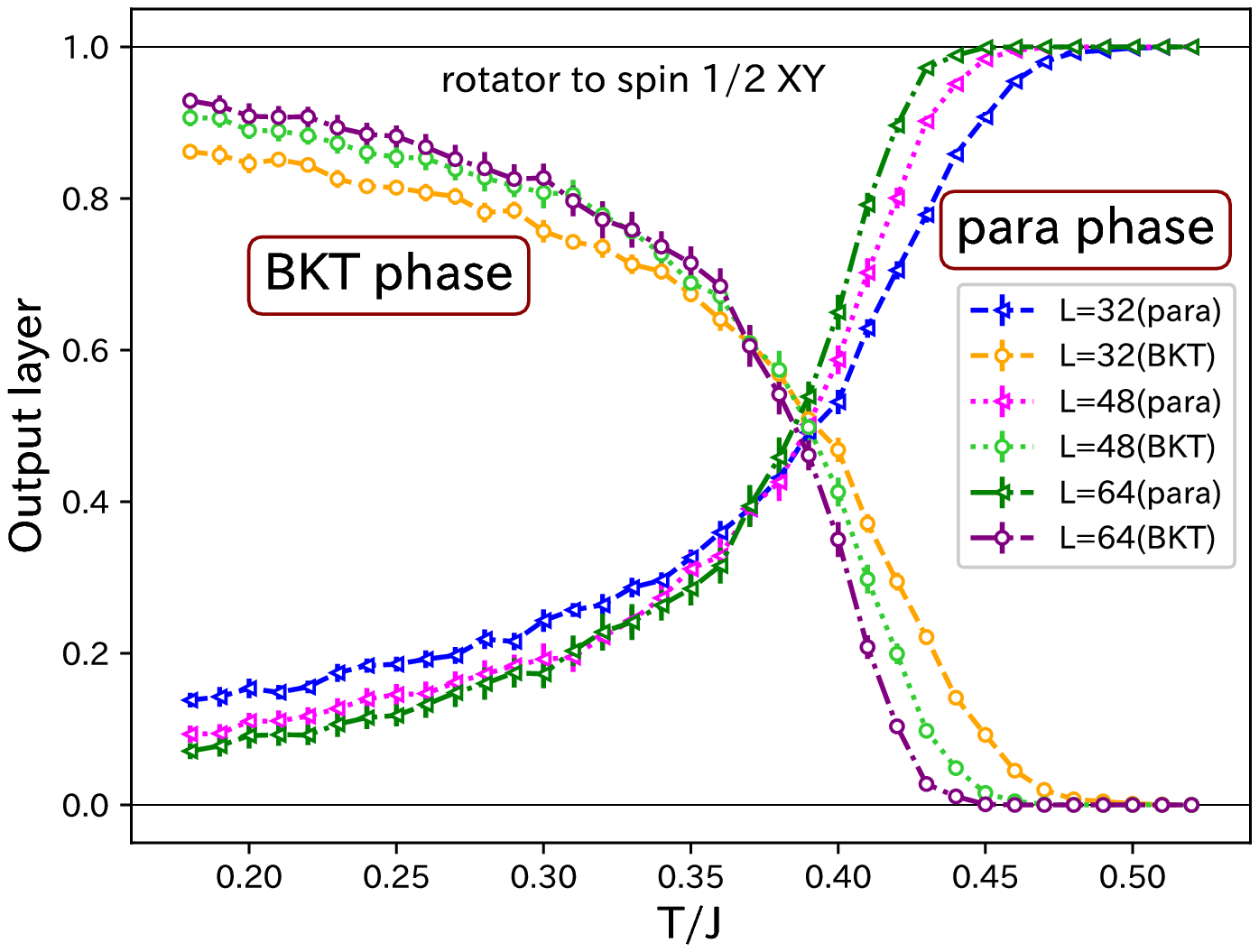}
\caption{
(a) The output layer averaged over a test set as a function of $T$ 
for the 2D spin 1/2 XY model. The system sizes are $L$ = 32, 48, and 64. 
The samples of $T$ within the ranges $0.16 \le T \le 0.32$ and 
$0.38 \le T \le 0.54$ are used for the training data. 
(b) The classification of the quantum XY model using the 
training data of the classical XY model (plane rotator). 
The samples of $T$ within the ranges $0.50 \le T \le 0.84$ and 
$0.96 \le T \le 1.30$ for the classical XY model are used 
for the training data. 
}
\label{fig:quantum}
\end{figure}

The classification of the BKT and paramagnetic phases 
of the spin 1/2 XY model using the machine-learning technique 
is shown in Fig.~\ref{fig:quantum}(a). 
The samples of $T$ within the ranges $0.16 \le T \le 0.32$ 
and $0.38 \le T \le 0.54$ were used for the BKT-temperature 
and high-temperature training data, respectively. 
If we estimate the value of $T_{\mathrm{BKT}}$ from the point 
that the probabilities of predicting two phases are 50\%, 
this temperature becomes around $T=0.40$. 
It is slightly higher than the precise estimate 
for the infinite system, $T_{\mathrm{BKT}}=0.342$~\cite{Harada}, 
although this temperature becomes lower 
as the system size increases. 
We also tested the classification of the quantum XY model 
using the training data of the classical model. 
For the classical model, not only the classical XY model 
(plane rotator) but also the anisotropic Heisenberg model 
with XY interaction was treated. This anisotropic Heisenberg 
model has out-of-plane fluctuation and the BKT transition 
temperature is slightly lowered at around 
$T_{\mathrm{BKT}}=0.70$~\cite{Evertz96,Figueiredo}. 
In Fig.~\ref{fig:quantum}(b), we show the result of 
the classification of the quantum XY model using the 
training data of the classical XY model (plane rotator). 
We reproduced the BKT transition of the quantum XY model. 
The same conclusion was obtained when using the anisotropic 
Heisenberg model as the training data. 
The classification into two phases is slightly sharper for the anisotropic 
Heisenberg model than the classical XY model (plane rotator). 
The opposite direction, using the training data of the quantum model 
in the classification of the classical models, was also successful.

To summarize, we have proposed a method to use the improved estimator 
of the correlation configuration in the machine-learning study 
of the phase classification of spin models.  
For the classical spin systems, we have demonstrated 
the machine-learning studies of the 2D 3-state Potts model 
(the second-order transition) and the 2D 6-state clock model 
(the BKT transition).  The results were compared with 
those of the previous study~\cite{Shiina} using 
the ordinary correlation instead of the improved correlation. 
The method was also applied to the quantum Monte Carlo simulation 
using the loop algorithm. 
We treated the spin 1/2 quantum XY model, 
and analyzed the BKT transition of the model.
We emphasize that the classification scheme based on the 
training data of the classical XY model can be used 
for the phase classification of the quantum model. 
It indicates the universality of the phase transition, 
and at the same time, the generalized feature of 
the phase classification based on the 
machine learning.  We also point out the effectiveness 
of the improved estimators in the loop algorithm 
to bridge classical and quantum Monte Carlo simulations.

We have opened a door to using the improved estimators 
for the machine-learning study of quantum systems. 
It is not trivial whether loop clusters in
quantum spin systems can be identified with
FK clusters in classical spin systems~\cite{nonomura2020}.
In this study, we clarified that
the phase classification using machine learning
does not discriminate between loop clusters and FK clusters. 
The BKT transition of the present study is a thermal phase transition. 
The investigation of a quantum phase transition at $T=0$ 
will be interesting. 
For future studies, we may list up several models 
for spin and charge degrees of freedom with loop algorithms. 
Examples are several quantum spin models, strongly-correlated 
electron models, hard-core boson models, optical lattices, etc.

Another direction of the future study is related to 
the inverse renormalization group approach~\cite{Ron}. 
Efthymiou \textit{et al.}~\cite{Efthymiou} have proposed a method 
to increase the size of lattice spin configuration using 
super-resolution, deep convolutional neural networks. 
At high temperatures, however, there is a problem that 
the noise is largely random and difficult to learn. 
The present improved correlation configuration could reduce 
this difficulty 
at high temperatures.

\vspace*{3mm}
The authors thank Hiroyuki Mori for valuable discussions. 
This work was supported by a Grant-in-Aid for Scientific Research 
from the Japan Society for the Promotion of Science Grant Number JP16K05480, 
JP16K05482.
KS is grateful to the A*STAR (Agency for Science, Technology and Research) 
Research Attachment Programme (ARAP) of Singapore 
for financial support. 

\bibliography{Tomita}

\end{document}